# Wave Scattering at temporal interfaces with spatial-translation-symmetry mismatch

Cong Chen[1], Kaijun Yi[1,*], Guoliang Huang[2], Gengkai Hu[1,3,*]

[1]School of Aerospace Engineering, Beijing Institute of Technology, Beijing 100081, China

[2] Department of Mechanics and Engineering Science, College of Engineering, Peking University, Beijing 100871, China.

[3] Marine Science and Technology Domain, Beijing Institute of Technology (Zhuhai), Zhuhai 519088, China

*Corresponding author: kaijun.yi@bit.edu.cn (K.Y.), hugeng@bit.edu.cn (G.H.)

**Abstract:**

Temporal interfaces enable wave manipulation through broken time-translation symmetry, but conventional formulations generally assume that spatial-translation symmetry is preserved across the interface. Here we consider temporal interfaces between periodic media with mismatched spatial symmetries. It is discovered that the reciprocal-lattice vectors of the pre- and post-switching media enter a generalized quasi-momentum-matching condition, giving rise to reciprocal-lattice-assisted wave-vector conversion. We then develop a multichannel temporal-scattering theory and validate it in one- and two-dimensional elastic lattices. A single incident Bloch mode can thereby excite multiple post-interface Bloch modes with distinct wave vectors and frequencies, a response inaccessible at conventional temporal interfaces. These results establish symmetry mismatch as a new degree of freedom for simultaneous control of wave vector and frequency in time-modulated periodic media.

*Introduction.* Temporal interfaces, at which material parameters are abruptly changed in time [1], provide a powerful way to control waves beyond the limits of static media. By breaking continuous time-translation symmetry [2] they allow waves to exchange energy with the time-varying medium, leading to temporal reflection and refraction [3-11]. When further combined with resonant dispersion, anisotropy, bianisotropy, non-Hermitian and other material responses, this principle has enabled a broad range of temporal wave phenomena, including frequency splitting [12,13], temporal evanescent modes [14], temporal aiming [15], mode conversion [16,17], polarization-dependent scattering [18,19] and topological temporal boundary states [20].

Despite these advances, existing temporal-interface physics remains constrained by the conventional assumption that spatial-translation symmetry is identical on both sides of the temporal interface. In a spatially homogeneous medium [Fig. 1(a1)], this symmetry forbids coupling between plane waves with different wave vectors. Thus, a conventional temporal interface can change the frequency of a wave, but its wave vector is conserved, $\mathbf{k}' = \mathbf{k}$ [Fig. 1(a2)][21-28]. The same restriction persists in periodic media with unchanged lattice periodicity. In such media, the eigenmodes are Bloch waves labeled by a Bloch wave vector, defined modulo reciprocal-lattice vectors. Previous studies have primarily considered temporal interfaces between crystals with the same lattice periodicity [Fig. 1(b1)], for which the reciprocal lattice is identical on both sides of the interface. Temporal scattering then obeys the usual Bloch-momentum selection rule: $\mathbf{k}' = \mathbf{k}$ within the first Brillouin zone [Fig. 1(b2)].

Here, we remove this constraint by considering temporal interfaces between periodic media with mismatched spatial translation symmetries. The lattices before and after the temporal interface have different periods and, consequently, different reciprocal-lattice vectors. In this case, no single Bloch wave vector is conserved across the interface. Instead, temporal scattering must be described by a generalized matching rule for quasi-momentum, taking into account the different reciprocal lattices on the two sides of the interface. This rule enables reciprocal-lattice-assisted coupling between Bloch modes with different wave vectors, opening temporal scattering channels that are forbidden when the spatial symmetry is preserved. We develop a multichannel temporal-scattering framework and apply it to elastic lattices, revealing that symmetry mismatch can induce simultaneous wave-vector and frequency conversion, together with controlled wave scattering. The proposed mechanism can be applicable to other wave systems, including acoustic, optical, and electromagnetic platforms, and establishes a foundation for selectively manipulating temporal-scattering channels with prescribed wave vectors.

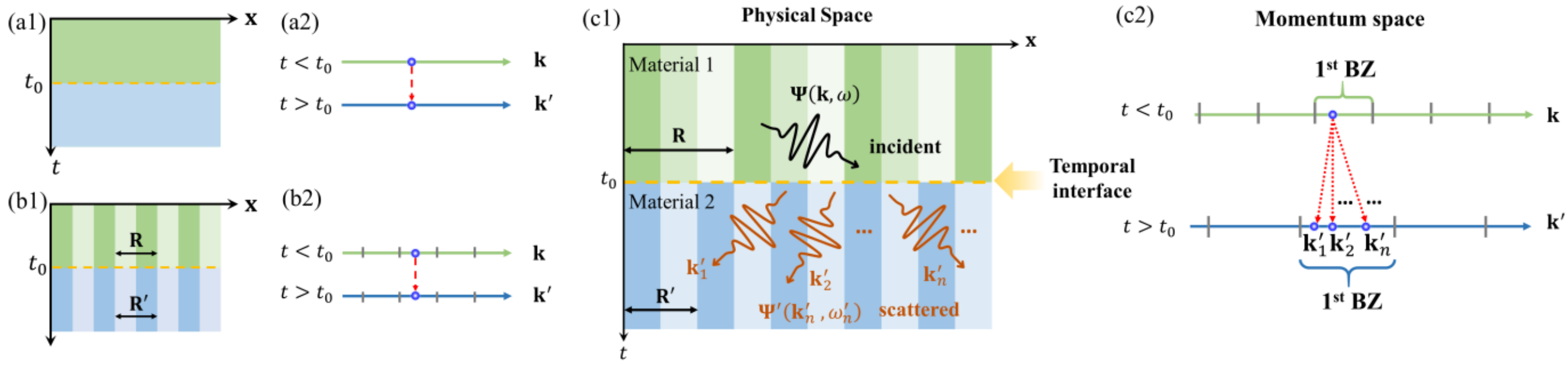

Fig. 1. Different types of temporal interfaces in physical space and corresponding wave-vector mapping in momentum space. (a1),(a2) Temporal interface between two homogeneous continuous media, where continuous spatial translation symmetry enforces strict momentum conservation. (b1),(b2) Temporal interface between periodic media with unchanged lattice translations $\mathbf{R} = \mathbf{R}'$, where the Bloch wave vector is conserved within the first Brillouin zone (1st BZ). (c1),(c2) Wave scattering at a temporal interface between periodic media with mismatched lattice translations $\mathbf{R} \neq \mathbf{R}'$ where quasi-momentum conservation leads to one-to-multiple Bloch wave-vector conversion within the 1st BZ. $\mathbf{\Psi}$ and $\mathbf{\Psi}'$ denote the wave fields before and after the interface, respectively.

*Generalized quasi-momentum conservation*. Consider a Bloch wave $\mathbf{\Psi}(\mathbf{x}, t)$ propagating in material 1 before the temporal interface $(t < t_0)$ [Fig. 1(c1)]. Expanding its periodic envelope into reciprocal-lattice harmonics, the field can be written as $\mathbf{\Psi}(\mathbf{x}, t) = \sum_{\mathbf{G}} C_{\mathbf{G}} e^{i\omega t - i(\mathbf{G}+\mathbf{k})\mathbf{x}}$, where $\mathbf{G}$ runs over all reciprocal-lattice vectors of material 1. At $t = t_0$, the material parameters are instantaneously changed, switching the lattice vectors from $\mathbf{R}$ to $\mathbf{R}'$ and thereby creating a mismatch between the pre- and post-interface spatial translation symmetries. The time-scattered field $\mathbf{\Psi}'(\mathbf{x}, t)$ in material 2 is then expanded over all allowed post-interface Bloch channels, characterized by amplitude $A'_n$, frequency $\omega'_n$ and Bloch wave vector $\mathbf{k}'_n$. Since each post-interface mode follows the periodicity of material 2, it can be decomposed into Fourier harmonics over the reciprocal lattice $\mathbf{G}'$, giving $\mathbf{\Psi}'(\mathbf{x}, t) = \sum_n A'_n \sum_{\mathbf{G}'} C_{\mathbf{G}',n} e^{i\omega'_n t - i(\mathbf{G}'+\mathbf{k}'_n)\mathbf{x}}$. Imposing wave-field continuity at the temporal interface, $\mathbf{\Psi}(\mathbf{x}, t)|_{t=t_0^-} = \mathbf{\Psi}'(\mathbf{x}, t)|_{t=t_0^+}$, yields

$$\sum_{\mathbf{G}} C_{\mathbf{G}} e^{i\omega t_0 - i(\mathbf{G}+\mathbf{k})\mathbf{x}} = \sum_n A'_n \sum_{\mathbf{G}'} C_{\mathbf{G}',n} e^{i\omega'_n t_0 - i(\mathbf{G}'+\mathbf{k}'_n)\mathbf{x}} . \tag{1}$$

Eq. (1) holds for arbitrary $\mathbf{x}$, the spatial phase factors of harmonics on both sides of the equation must be matched. Therefore, a quasi-momentum conservation condition is obtained at the temporal interface

$$\mathbf{G} + \mathbf{k} = \mathbf{G}' + \mathbf{k}'_n . \tag{2}$$

Eq. (2) shows that relaxing the constraint of unchanged spatial translation symmetry allows reciprocal-lattice vectors on both sides to participate in the momentum balance across the temporal interface. This generalized matching condition therefore provides the basis for reciprocal-lattice-assisted wave-vector conversion across the interface.

Although similar relations appear in phonon Umklapp processes [29] and interlayer interaction in bilayer systems [30-33], Eq. (2) plays a distinct role in temporal scattering. Here, it serves as a reciprocal-lattice-assisted selection rule for the allowed converted wave-vector channels. This mechanism can be interpreted as a generalized Umklapp process at a temporal interface, as the temporal counterpart of conventional spatial Umklapp scattering [29-32]. The widely studied case of unchanged lattice periodicity [34-38] can be treated as a special case: when the reciprocal lattices before and after the temporal interface coincide, Eq. (2) reduces to conservation of the Bloch wave vector within the first Brillouin zone.

*Multichannel temporal-scattering theory.* The selection rule in Eq. (2) may lead to a one-to-multiple wave-vector mapping when the spatial translation symmetries are mismatched [Fig. 1(c2)]. This mapping originates from the Fourier decomposition of the incident Bloch mode, whose spatial harmonics $e^{-i(\mathbf{k}+\mathbf{G})\mathbf{x}}$ carry wave vectors $\mathbf{k}+\mathbf{G}$. Although these harmonics are equivalent to the same Bloch wave vector $\mathbf{k}$ in material 1, a change in spatial periodicity makes the post-interface reciprocal lattice different. Consequently, different admissible harmonics can be folded back into the first Brillouin zone of material 2 modulo its reciprocal-lattice vectors $\mathbf{G}'$, yielding distinct mapped Bloch wave vectors.

To determine the scattered waves analytically, we develop a multichannel temporal-scattering theory by first identifying the allowed wave-vector channels and their number. For clarity, we begin by considering commensurate systems, where the pre- and post-interface lattices share a minimal common supercell with lattice vector $\mathbf{R}_c$ and reciprocal vectors $\mathbf{G}_c$. Because the translational symmetry of this supercell is preserved across the temporal interface, the incident and scattered Bloch waves must acquire the same phase under translations by $\mathbf{R}_c$, i.e., $e^{-i\mathbf{k}\mathbf{R}_c}=e^{-i\mathbf{k}'\mathbf{R}_c}$. This condition determines how a single incident wave vector $\mathbf{k}$ maps to multiple post-interface wave vectors $\mathbf{k}'$. First, the pre-interface lattice is described using the common supercell. In this representation, the incident wave has a supercell wave vector $\mathbf{k}_c$, obtained by folding $\mathbf{k}$ into the first supercell Brillouin zone if necessary. The temporal interface is then treated in the same supercell basis. Since the supercell translation symmetry remains preserved, all scattered waves retain the same $\mathbf{k}_c$, while their frequencies are set by the post-interface band structure. Finally, these scattered waves are rewritten in the primitive Bloch basis of the post-interface lattice. Let $V_r'$ and $V_r^c$ be the reciprocal-space volumes of the primitive post-interface Brillouin zone and the common-supercell Brillouin zone, respectively. Their ratio $N_k = \frac{V_r'}{V_r^c}$ is an integer, that counts the number of supercell Brillouin zones contained in the primitive Brillouin zone of material 2. Unfolding $\mathbf{k}_c$ into these $N_k$ zones and mapping the resulting vectors into the first Brillouin zone of the post-interface lattice gives $N_k$ distinct wave vectors $\mathbf{k}' = \mathbf{k}_c + \mathbf{G}_c$. This procedure is illustrated in Fig. 2 for the one-dimensional lattice considered below.

Once the mapped wave vectors $\mathbf{k}'$ are obtained, the temporal scattering process can be solved systematically. For each mapped $\mathbf{k}'$, the post-interface dispersion relation gives the positive-frequency branches and their negative-frequency counterparts, corresponding to the time-refracted and time-reflected Bloch modes. These amplitudes are then determined by imposing the temporal boundary conditions on a superposition of post-interface Bloch modes selected by the mismatched reciprocal lattices [see Supplemental Material for details].

The above procedure can also be extended to incommensurate systems. In such cases, no finite common supercell exists, which implies that an infinite number of Bloch vectors are allowed after the interface as $V_r^c \to 0$. Nevertheless, the problem can be solved using a sufficiently large finite supercell that captures the dominant temporal-scattering channels [Supplemental Material I].

Next, we apply the above theory to elastic systems and validate it using two representative examples in one- and two-dimensional elastic lattices, demonstrating reciprocal-lattice-assisted wave-vector conversion and the associated wave phenomena.

*Elastic-Lattice Examples*. We first consider a one-dimensional elastic lattice to illustrate wave-vector conversion. Before the temporal interface at $t_0$, the primitive cell [unit cell 1 in Fig. 2(a)] contains three identical masses $m = 1$ kg, connected by springs with $K_1 = 1$, $K_2 = 2$, $K_3 = 3$ N/m. At $t = t_0$, the lattice is abruptly changed to another periodic configuration, whose primitive cell (unit cell 2) contains two masses connected by springs with $K_1' = 1$, and $K_2' = 2$ N/m. We consider an incident wave with angular frequency $\omega_0 = 0.250$ rad/s, corresponding to a Bloch wave vector $k_0 = 0.062\pi/a$ in the lattice before $t_0$ [Fig. 2(b1)], where $a$ is the lattice constant. In the supercell representation, it is characterized by the supercell quasi-momentum $k_c = k_0 = 0.062\pi/a$ [Fig. 2(b2)], which is conserved after the interface [Fig. 2(b3)]. When represented in the first Brillouin zone of unit cell 2 [Fig. 2(b4)], $k_c$ unfolds into three inequivalent Bloch wave vectors $k' = k_c + lB_c$, with $l = 0, -1, 1$ and $B_c = \pi/3a$, giving $k_1' = 0.062\pi/a$, $k_2' = -0.271\pi/a$, and $k_3' = 0.396\pi/a$. Each mapped wave vector supports two positive-frequency branches, yielding six positive-frequency Bloch modes [Fig. 2(b4)], each accompanied by a negative-frequency counterpart. The normalized scattering amplitudes of the positive-and negative-frequency Bloch modes are obtained as

$$|F_{j,m}| = \left|\frac{1}{2}\left(1 + \frac{\omega}{\omega_{j,m}}\right)\boldsymbol{\varphi}_{j,m}^{\dagger}\boldsymbol{\varphi}_0\right| \tag{3}$$

$$|B_{j,m}| = \left|\frac{1}{2}\left(1 - \frac{\omega}{\omega_{j,m}}\right)\boldsymbol{\varphi}_{j,m}^{\dagger}\boldsymbol{\varphi}_0\right| \tag{4}$$

Here, $\boldsymbol{\varphi}_0$ is the normalized incident eigenvector, $\boldsymbol{\varphi}_{j,m}$ is the normalized eigenvector associated with the $m$-th positive-frequency branch at the mapped wave vector $k_j'$, and $\omega_{j,m}$ is the corresponding frequency. In contrast to the expressions for conventional temporal interfaces with unchanged periodicity [14,39], $\boldsymbol{\varphi}_0$ and $\boldsymbol{\varphi}_{j,m}$ are defined in the common-supercell basis. Therefore, Eqs. (3) and (4) provides the basis for energy analysis among scattering channels with different wave vector and frequencies. Further details are provided in Supplemental Material II.

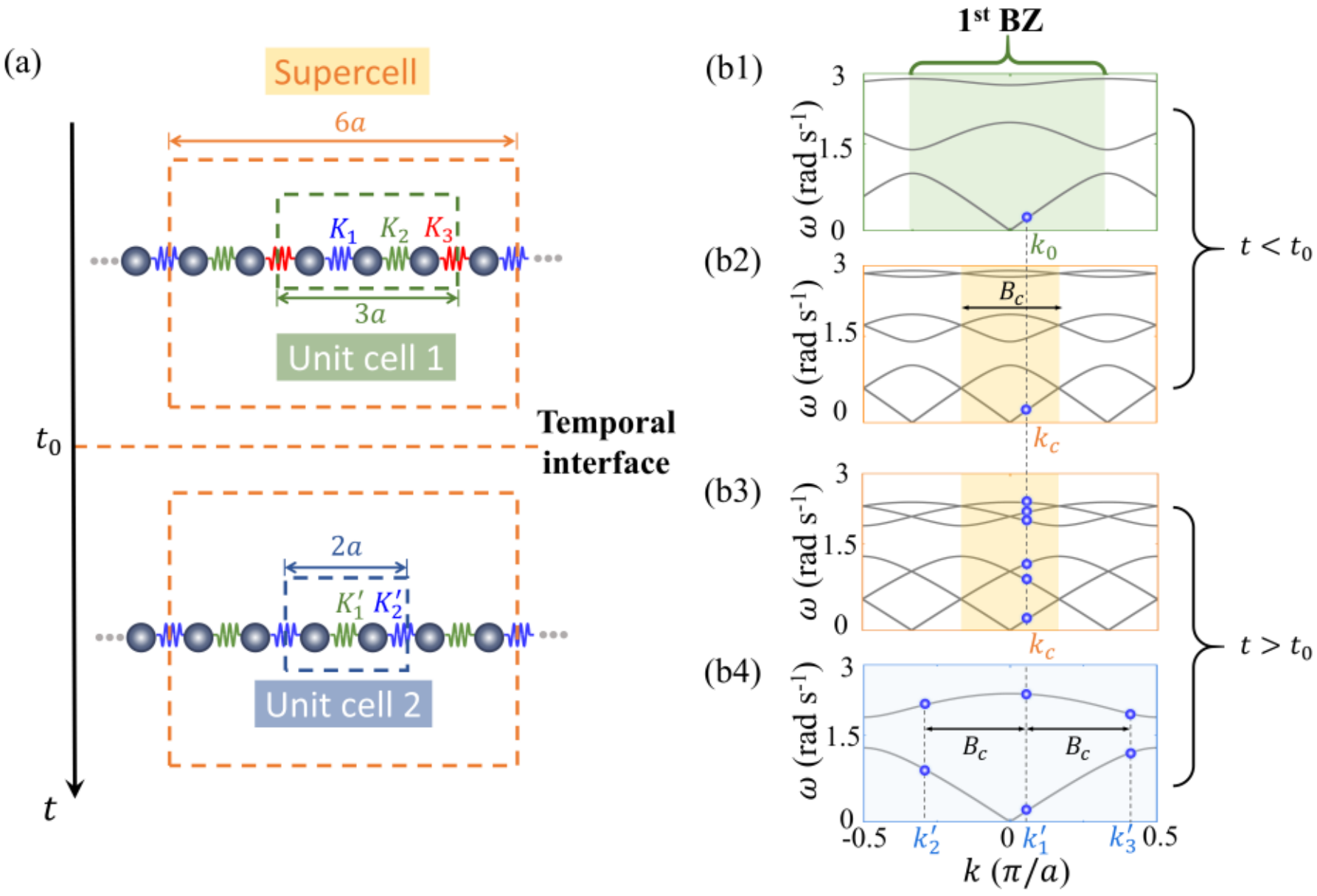

Fig. 2 Temporal interface and wave-vector conversion in a one-dimensional elastic lattice. (a) Triatomic-to-diatomic temporal interface with $m = 1$ kg, $K_1 = 1$, $K_2 = 2$, $K_3 = 3$ N/m, in unit cell 1, and $K_1' = 1$, $K_2' = 2$ N/m in unit cell 2. The supercell contains six masses. (b) Wave-vector conversion across the interface. Panels (b1),(b2) show the pre-interface

dispersion relations in the unit cell 1 and supercell representations, respectively, (b3),(b4) show the corresponding post-interface dispersion relations. For clarity, only the positive-frequency branches are plotted, and each branch has a corresponding negative-frequency counterpart. Here, $B_c = \pi/3a$ denotes the reciprocal basis vector of the supercell. The incident Bloch wave vector $k_0$ folds to $k_c$ in the 1st BZ of supercell, and then unfolds into three Bloch wave vectors $k_1'$, $k_2'$ and $k_3'$ in the 1st BZ after $t_0$.

To validate simultaneous frequency and wave-vector conversion, we perform finite-difference time-domain (FDTD) simulations of the temporal interface shown in Fig. 2(a). A wave packet is launched from the left side of a finite lattice containing 4500 masses. The simulation runs from $t = 0$ to $t_1 = 70T$, with $T = 2\pi/\omega_0$, and the spring constants are abruptly switched at $t_0 = 35T$ when the packet reaches the center of the structure. As illustrated in Fig. 3(a), the incident wave packet splits into multiple pairs of propagating waves after $t_0$. To further visualize the results, spatial Fourier analysis is performed over different spatial regions of the displacement field snapshot at $t_1 = 70T$ [Fig. 3(b)]. The spectra reveal 12 distinct waves in total, with three waves in each of regions (1)–(4). Their wavenumbers and amplitudes agree well with the theoretical predictions obtained from Eqs. (3) and (4), marked by circles. To verify the associated frequency conversion, temporal Fourier analysis of the displacement at mass index $i = 2400$ over interval $t_0 \leq t \leq t_1$ further reveals six spectral peaks [Fig. 3(c)], all in excellent agreement with the predicted frequencies marked by vertical dashed lines.

The energy partition among different wave-vector channels is shown in [Fig. 3(d1)]. In this example, the channel for $k_1' = k$ carries $\eta(k_1') =$ 90.32% of the total energy after the interface. Importantly, for a given incident wave vector, the energy partition among the scattered channels can be tuned by changing the material parameters and modifying the modal overlap. As a result, most of the energy can be transferred to the scattering channels with converted wave vector $k_2'$ and $k_3'$. For example, when the stiffness parameters are chosen as $K_1 = 1$, $K_2 = 0.1$, $K_3 = 5$, $K_1' = 1$, and $K_2' = 5$ N/m, $\eta(k_1')$ is reduced to 43.15 % [Fig. 3(d2)]. The minimum value of $\eta(k_1')$ is 33.33% [Fig. 3(d3)], corresponding to an equal energy partition among the three wave-vector scattering channels. We expect this constraint can be further relaxed by introducing negative stiffness. Based on these results, energy distribution can be tuned and controlled among different scattering channels.

To illustrate the generality of the framework, a one-dimensional incommensurate example is also provided in Supplemental Material I, where a temporal interface connects two continuous media whose spatial modulation periods have an irrational ratio.

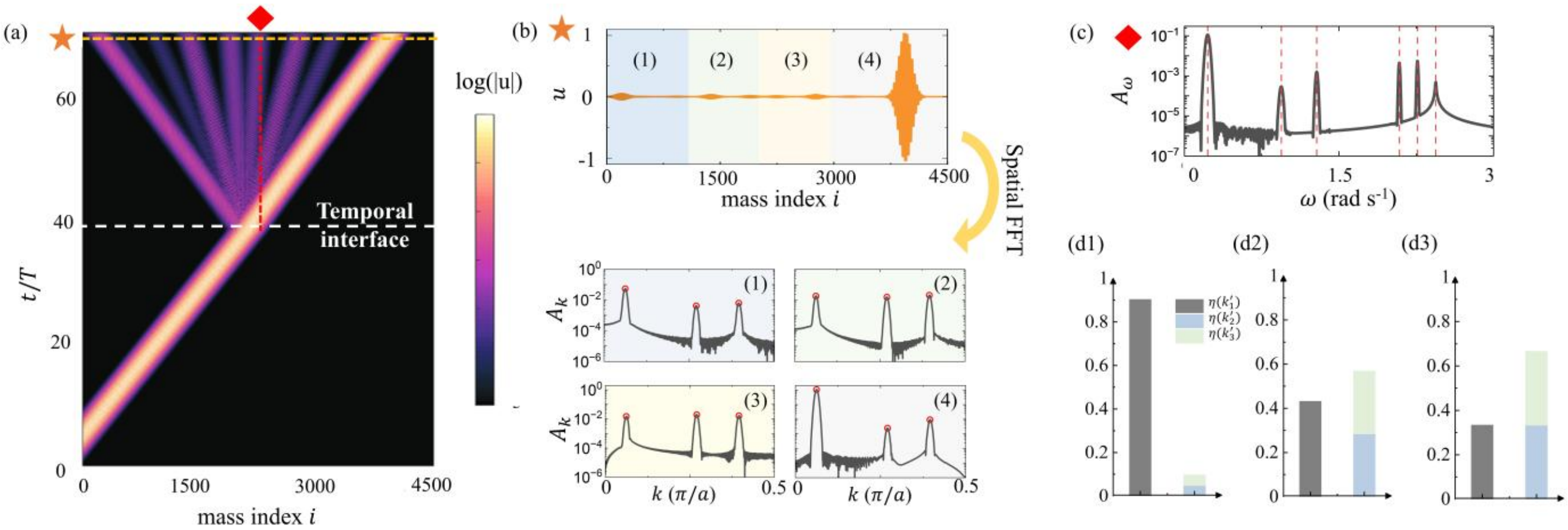

Fig. 3 Results for the one-dimensional elastic lattice. (a) Displacement field evolution shown in logarithmic scale, normalized by the amplitude of the incident wave. (b) Snapshot of the displacement field at $t_1 = 70T$, together with the corresponding spatial Fourier spectra obtained from regions (1)–(4). These regions correspond to the mass-index ranges 1–1150, 1151–2050, 2051–3000, and 3001–4500, respectively. The Fourier spectra are displayed on a logarithmic scale, and the red circles denote the theoretically predicted wavenumbers and amplitudes. (c) Temporal Fourier spectrum of the displacement at mass index $i = 2400$ computed over the time interval $(t_0, t_1)$. The red vertical dashed lines indicate the theoretically predicted frequencies. (d) Energy fractions of the scattered waves in different wave-vector channels. Each wave-vector channel contains four waves, corresponding to two positive-frequency and two negative-frequency components. The stiffness parameters are set as (d2) $K_1 = 1$, $K_2 = 0.1$, $K_3 = 5$, $K_1' = 1$, and $K_2' = 5$ N/m; (d3) $K_1 = 1$, $K_2 = 2\times10^{-4}$, $K_3 = 200$, $K_1' = 1$, and $K_2' = 200$ N/m.

In higher-dimensional lattices, wave-vector conversion induced by mismatched spatial translation symmetries can further redirect wave propagation. To demonstrate this, we consider a two-dimensional elastic square lattice of identical masses $m = 1$ kg, connected by springs with alternating stiffness constants $K_{x1} = 1$, $K_{x2} = 2$ N/m in the horizontal direction and $K_{y1} = 1$, $K_{y2} = 2$ N/m in the vertical direction [Fig. 4(a)]. At $t = t_0$, the alternating stiffness pattern is abruptly replaced by a uniform distribution with $K_x' = 1$ and $K_y' = 1$ N/m, reducing the post-interface primitive cell from four degrees of freedom to one. We consider an incident wave with angular frequency $\omega_0 = 1.000$ rad/s, incident angle $\theta = \frac{\pi}{4}$, and wave vector $\mathbf{k}_0 = (0.199, 0.199)\pi/a$. Here, the minimal common supercell is identical to the pre-interface primitive cell and contains four post-interface primitive cells. Consequently, $\mathbf{k}_0$ maps to four inequivalent post-interface wave vectors, $\mathbf{k}_1' = (0.199, 0.199)\pi/a$, $\mathbf{k}_2' = (-0.801, 0.199)\pi/a$, $\mathbf{k}_3' = (0.199, -0.801)\pi/a$, $\mathbf{k}_4' = (-0.801, -0.801)\pi/a$ [Fig. 4(b)(c)]. Each wave vector supports a positive-frequency branch and its negative-frequency counterpart, giving eight scattered wave packets in total. Group-velocity calculations further show that the packets associated with $\mathbf{k}_2'$ and $\mathbf{k}_3'$ are deflected away from the incident direction, demonstrating temporal-interface-induced wave-packet redirection [Supplemental Material III].

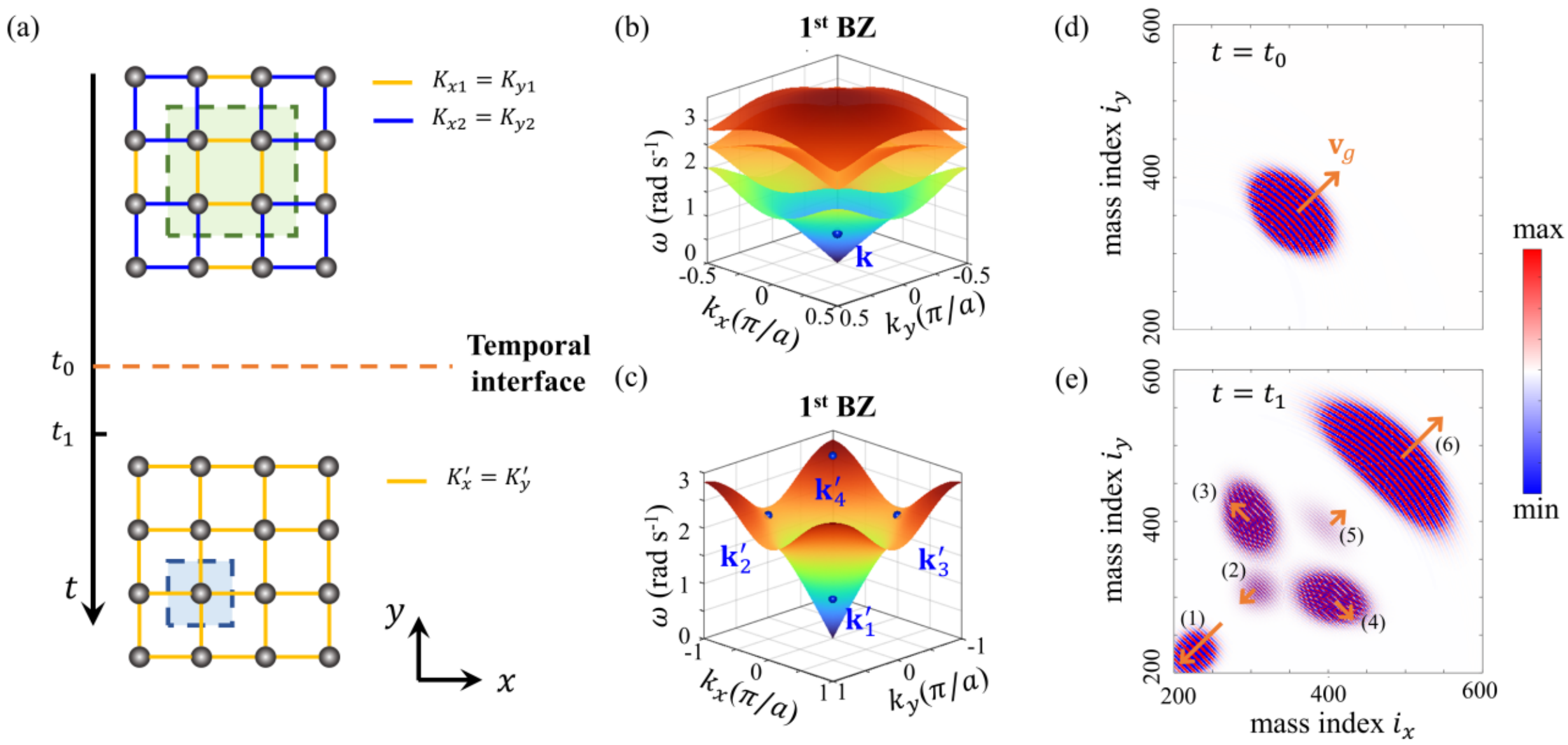

Fig. 4 Two-dimensional elastic-lattice example. (a) Temporal interface between a lattice with alternating stiffnesses $K_{x1} = 1$, $K_{x2} = 2$, $K_{y1} = 1$, and $K_{y2} = 2$ N/m, and a uniform post-interface lattice with $K_x' = 1$ and $K_y' = 1$ N/m. All masses are $m = 1$ kg. The green and blue dashed cells are the primitive cells before and after the interface, respectively. (b),(c) Wave-vector mapping in the 1st BZ across the temporal interface. (d),(e) Out-of-plane Displacement fields at $t = t_0$ and $t = t_1$, shown over mass indices 200–600 along both x and y directions. Arrows denote group velocities of the wave packets, with their directions and lengths denoting the propagation directions and relative magnitudes, respectively. Six spatially resolved wave packets are labeled (1)–(6) in (e), and packets (3) and (4) each contain two overlapping sub-packets with identical group velocities.

To validate the theoretical predictions, we perform FDTD simulations from $t = 0$ to $t_1 = 70T$, where $T = 2\pi/\omega_0$. The temporal interface is introduced at $t_0 = 40T$ and the simulation domain contains 600×600 masses. Fig. 4(d),(e) present the spatial wave fields at $t_0$ and $t_1$, respectively. After the temporal interface, the incident wave packet is converted into several scattered packets propagating along distinct directions. Six wave packets are spatially resolved, fewer than the eight predicted components, because packets (3) and (4) each consist of two overlapping sub-packets with identical group velocities, which can be distinguished by spatial Fourier analysis [Supplemental Material III]. The extracted wave vectors, frequencies, and amplitudes agree well with the theoretical predictions, as detailed in Supplemental Material III. These results show that a temporal interface with mismatched spatial translation symmetries can convert one incident wave packet into multiple scattered packets with distinct propagation directions. This mechanism differs fundamentally from wave-packet redirection based on anisotropic time-varying materials [15,40]: here, the redirection arises from the mismatched spatial translation symmetries across the temporal interface and is accompanied by multiple wave-vector conversion.

In this example, the lattice-vector directions remain unchanged, so the wave-vector mapping can be interpreted as the combination of independent mappings along the $x$ and $y$ directions. Cases involving changes in the directions of the primitive lattice vectors are also covered by our framework, as provided in Supplemental Material IV.

In summary, we have investigated wave scattering at temporal interfaces between periodic media with mismatched spatial-translation symmetries. Unlike conventional temporal interfaces, which preserve spatial-translation symmetry and hence wave vector, these interfaces obey a generalized quasi-momentum conservation condition involving reciprocal-lattice vectors from both the initial and final media. This condition enables reciprocal-lattice-assisted wave-vector conversion, providing a temporal analogue of Umklapp scattering. Consequently, a single incident Bloch mode can excite multiple post-interface modes with distinct wave vectors and frequencies, introducing a new degree of freedom for their simultaneous control. We formulate a multichannel temporal-scattering theory and validate the predicted frequency–wave-vector conversion and wave-packet redirection in elastic lattices. These results provide a basis for inverse-designed temporal scattering in which lattice symmetries are engineered to route waves into prescribed frequency and wave-vector channels. For physical realization, the proposed symmetry-mismatched temporal interfaces could be implemented in mechanical metamaterials with piezoelectric shunt circuits [11,41,42]. This reciprocal-lattice-assisted mechanism is also applicable to incommensurate systems through supercell approximations and can be extended to acoustic, mechanical, and photonic crystals, enabling joint control of frequency and wave vector at temporal interfaces.